\documentclass[12pt]{article}
\usepackage{graphicx}
\usepackage{xcolor}
\usepackage{subcaption}
\usepackage{amsmath}
\usepackage{multirow}
\usepackage{slashed}

\def\as{\alpha_s}

\newcommand\pubnumber{TTP-22-003\\ P3H-22-007}
\newcommand\pubdate{\today}

\def\institute{Institute for Theoretical Particle Physics\\ KIT, Karlsruhe, Germany}
\def\support{\footnote{Speaker}}

\def\Title#1{\begin{center} {\Large #1 } \end{center}}
\def\Author#1{\begin{center}{ \sc #1} \end{center}}
\def\Address#1{\begin{center}{ \it #1} \end{center}}

\newcommand\pubblock{\rightline{\begin{tabular}{l} \pubnumber\\
         \pubdate  \end{tabular}}}
\newenvironment{Abstract}{\begin{quotation}  }{\end{quotation}}
\newenvironment{Presented}{\begin{quotation} \begin{center} 
             PRESENTED AT\end{center}\bigskip 
      \begin{center}\begin{large}}{\end{large}\end{center} \end{quotation}}
\def\Acknowledgements{\bigskip  \bigskip \begin{center} \begin{large}
             \bf ACKNOWLEDGEMENTS \end{large}\end{center}}

\def\beq{\begin{equation}}
\def\eeq#1{\label{#1}\end{equation}}
\def\eeqn{\end{equation}}

\def\beqa{\begin{eqnarray}}
\def\eeqa#1{\label{#1}\end{eqnarray}}
\def\eeqan{\end{eqnarray}}

\let\bar=\overbar

\def\Dslash{\not{\hbox{\kern-4pt $D$}}}
\def\dslash{\not{\hbox{\kern-2pt $\del$}}}

\def\msb{{\bar{\ssstyle M \kern -1pt S}}}

\begin{document}
\begin{titlepage}
\pubblock

\vfill
\Title{Non-factorisable two-loop contribution to $t$-channel single-top production}
\vfill
\Author{Christian Br\o{}nnum-Hansen\support}
\Author{J\'{e}r\'{e}mie Quarroz}
\Author{Chen-Yu Wang}
\Address{\institute}
\vfill
\begin{Abstract}
We report on a recent computation of the non-factorisable contribution
  to the two-loop helicity amplitude for  $t$-channel single-top production~\cite{Bronnum-Hansen:2021pqc}. This is the last missing piece of the two-loop virtual corrections to this process. 
  We perform analytic integration-by-parts reduction to master integrals and use the auxiliary mass flow method for their fast numerical evaluation. The impact
  of the non-factorisable corrections to single-top production on experimentally measured observables is estimated.\end{Abstract}
\vfill
\begin{Presented}
$14^\mathrm{th}$ International Workshop on Top Quark Physics\\
(videoconference), 13--17 September, 2021
\end{Presented}
\vfill
\end{titlepage}
\def\thefootnote{\fnsymbol{footnote}}
\setcounter{footnote}{0}

\section{Introduction}

In these proceedings we give a brief summary of the calculation of  the non-factorisable contribution to the two-loop helicity amplitude for  $t$-channel single-top production presented recently in Ref.~\cite{Bronnum-Hansen:2021pqc}.

The heaviest particle in the Standard Model, the top quark, is frequently produced in hadron collisions at the LHC.
As its mass is entirely due to the Higgs field, studying top production can provide useful information about electroweak symmetry breaking within and beyond the Standard Model.

At the LHC, top and anti-top quarks are primarily produced in pairs through the strong interaction.
Alternatively, single (anti-)top quarks can be produced through charged electroweak interactions.
The cross section of single-top production is approximately a quarter of the $t \bar t$-pair cross section. This means that due to the high luminosity of the LHC, tens of millions of top quarks have been produced through this electroweak mechanism.

Single-top production offers many research opportunities, including the structure of the $tbW$ vertex~\cite{ATLAS:2017ygi,ATLAS:2019hhu}, CKM matrix elements~\cite{ATLAS:2019hhu,CMS:2020vac} and indirect  determination of the top quark width~\cite{CMS:2014mxl}.
Furthermore, top mass measurements from single-top events are becoming increasingly important~\cite{CMS:2017mpr}.

\begin{figure}[!h!tbp]
                \centering
                \begin{subfigure}[ht]{0.23\linewidth}
                    \centering
                    \includegraphics[height=3cm]{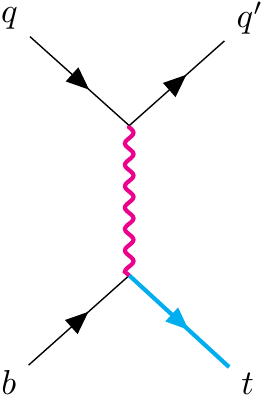}
                    \caption{Tree-level.}
                    \label{fig:tree}
                \end{subfigure}
                \hspace{2cm}
                \begin{subfigure}[ht]{0.23\linewidth}
                    \centering
                    \includegraphics[height=3cm]{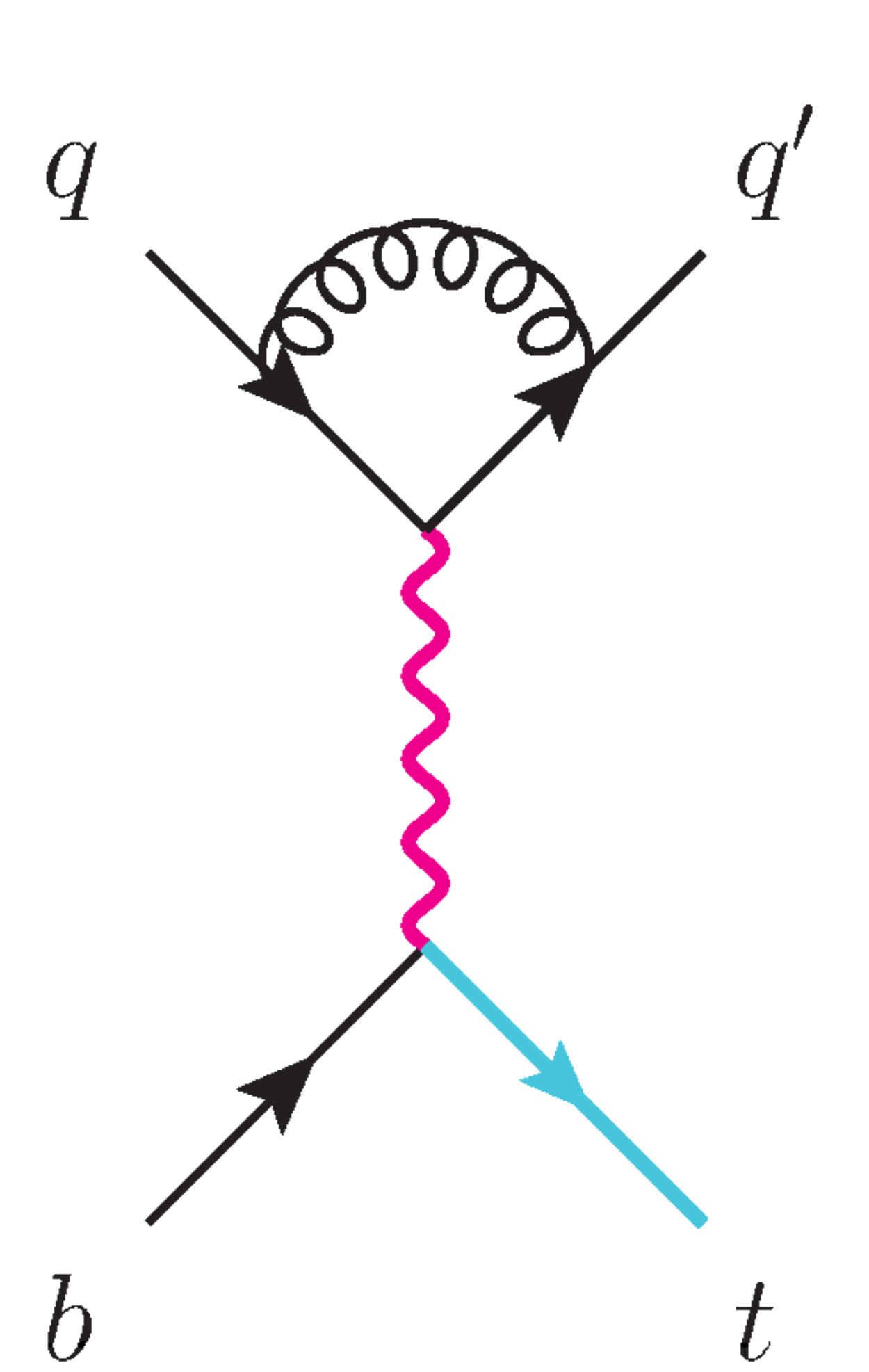}
                                        \caption{Factorisable.}
                    \label{fig:factorisable}
                \end{subfigure}
                \hspace{2cm}
                \begin{subfigure}[ht]{0.23\linewidth}
                    \centering
                    \includegraphics[height=3cm]{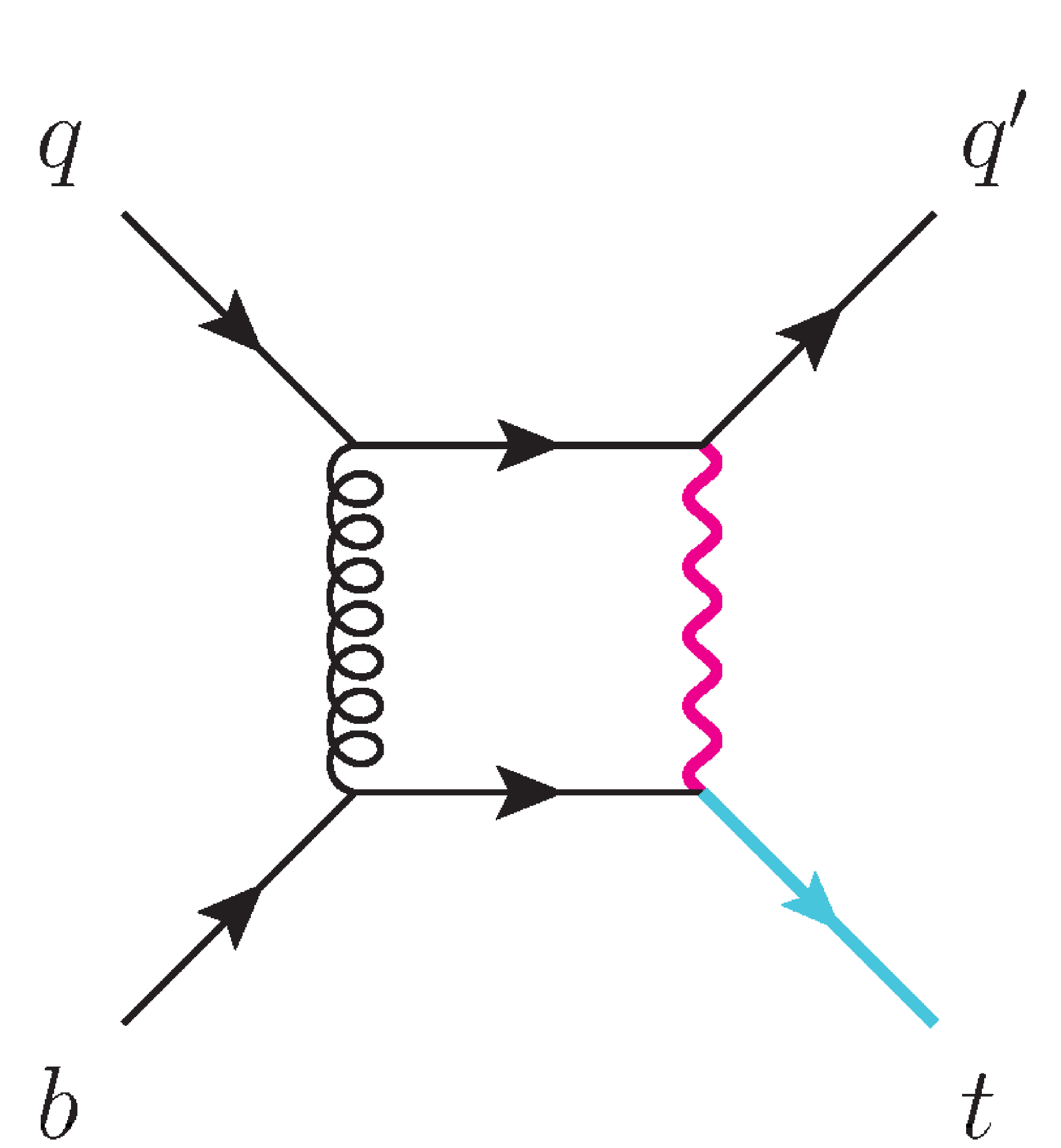}
                                        \caption{Non-factorisable.}
                    \label{fig:nonfactorisable-1l}
                \end{subfigure}
                \caption{Representative Feynman diagrams for $t$-channel single-top production at tree- and one-loop level. The pink, wavy line represents a $W$ boson.}
         \end{figure}

Single-top production proceeds primarily through the $t$-channel exchange of a $W$ boson~\cite{Giammanco:2017xyn}.
The only contributing diagram to this channel at leading order (LO) is given in Figure~\ref{fig:tree}.
At loop level we distinguish between factorisable, Figure~\ref{fig:factorisable}, and non-factorisable, Figure~\ref{fig:nonfactorisable-1l}, contributions.
The non-factorisable contributions are characterised by gluon exchanges between the quark lines.
These contributions are absent in the next-to-leading (NLO) cross section due to colour conservation.
At next-to-next-to-leading order (NNLO) non-factorisable two-loop diagrams are colour suppressed and have, for this very reason, been neglected in previous studies~\cite{Brucherseifer:2014ama,Berger:2016oht,Campbell:2020fhf}.
However, as was recently shown in the context of VBF Higgs production~\cite{Liu:2019tuy}, these contributions can be enhanced due to the so-called Glauber phase, largely compensating the colour suppression.

            \begin{figure}[!h!tbp]
                \centering
                \begin{subfigure}[ht]{0.44\linewidth}
                    \centering
                    \includegraphics[height=3cm]{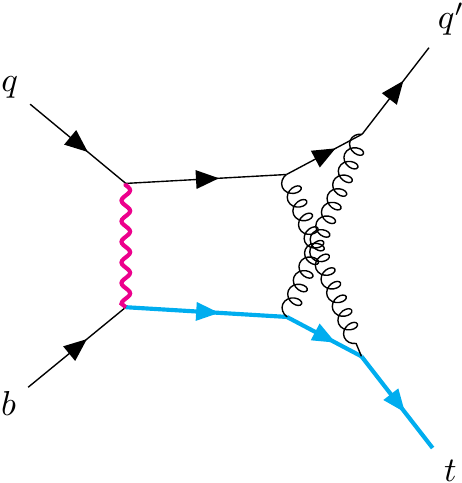}
                    \caption{Abelian vertices only.}
                    \label{fig:nonfactorisable-abelian}
                \end{subfigure}
                ~   
                \begin{subfigure}[ht]{0.44\linewidth}
                    \centering
                    \includegraphics[height=3cm]{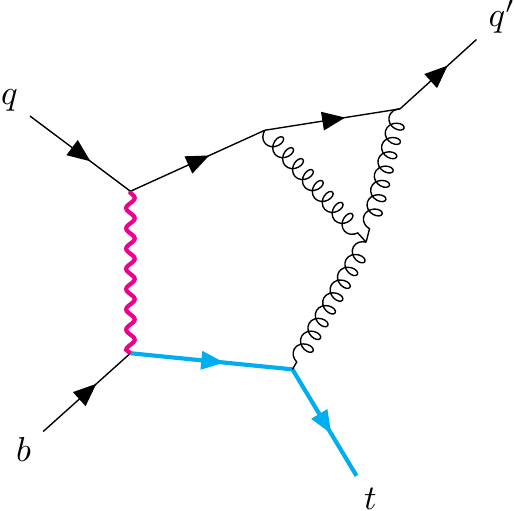}
                    \caption{Non-abelian three-gluon vertex.}
                    \label{fig:nonfactorisable-nonabelian}
                \end{subfigure}
                \caption{Examples of non-factorisable, two-loop diagrams for single-top production in the $t$-channel. Non-factorisable diagrams with non-abelian vertices do not contribute to the NNLO cross section due to colour conservation.}
            \label{fig:nonfactorisable-2l}
            \end{figure}

\section{Non-factorisable contributions to two-loop helicity amplitudes}\label{sec:ampcalc}

We study single-top production in the $t$-channel with the following assignment of flavours and momenta,
\begin{align}
  u(p_1) + b(p_2) &\to d(p_3) + t(p_4).
  \label{eq:process}
\end{align}
All external particles are considered to be on mass-shell such that $p_4^2 = m_t^2$ while $p_i^2 = 0$ for $i=1,2,3$. We consider the CKM matrix to be diagonal. 
Mandelstam variables are defined in a standard way, 
\begin{align}
s = (p_1 + p_2)^2,\qquad
t = (p_1 - p_3)^2,\qquad
u = (p_2 - p_3)^2,
\end{align}
with $ s+ t+u = m_t^2$.
We expand the amplitude
in the renormalised strong coupling constant  $\as = \as(\mu)$ as follows
\begin{align}
  \mathcal{A}(\{ p_i \}) =
 g_w^2 V_{ud} V_{tb} \left ( 
    \mathcal{A}^{(0)} + \frac{\as}{4\pi}
     \mathcal{A}_{\rm nf}^{(1)} + \left(\frac{\as}{4\pi}\right)^2
     \mathcal{A}_{\rm nf}^{(2)}+... + \mathcal{O}\left(\as^3\right)
     \right )
    \,.\label{eq:asexp}
\end{align}
We have extracted the weak coupling constants as well as two CKM matrix elements. We write only the non-factorisable amplitude up to two-loop order explicitly. Factorisable contributions are represented by ellipses.
At NNLO the cross section receives contributions from the interference between the two-loop scattering amplitude, $\mathcal{A}_{\rm nf}^{(2)}$, and the tree level diagram, $\mathcal{A}^{(0)}$, as well as from the one-loop amplitude, $\mathcal{A}_{\rm nf}^{(1)}$, squared.

Four box diagrams contribute to $\mathcal{A}_{\rm nf}^{(1)}$. Figure~\ref{fig:nonfactorisable-2l} shows examples of non-factorisable, two-loop diagrams.
All diagrams with non-abelian vertices, such as the one in Figure~\ref{fig:nonfactorisable-nonabelian}, or only a single gluon exchange between the quark lines vanish due to colour conservation upon interference with the LO amplitude.
Thanks to this colour conservation, only 18 double-box diagrams (planar and non-planar) contribute to $\mathcal{A}_{\rm nf}^{(2)}$.

We proceed by projecting the diagrams onto Lorentz invariant tensor structures.
At two loops we need 11 structures\footnote{Our tensor structures differ slightly from the ones used in Ref.~\cite{Assadsolimani:2014oga} and are given explicitly in Ref.~\cite{Bronnum-Hansen:2021pqc}.}.
The form factors are written as linear combinations of Feynman integrals with coefficients of rational functions in the kinematic invariants and masses.
We perform analytic integration-by-parts reduction\footnote{The very first reduction of the non-factorisable contributions to single-top production 
   to master integrals was performed using the approximation $m_t^2 = 14m_W^2/3$ in Ref.~\cite{Assadsolimani:2014oga}.} using Kira and Firefly~\cite{Klappert:2020nbg}.
This reduction allows us to express the two-loop amplitude in terms of 18 integral families with a total of 428 master integrals.

\section{Integral evaluation}

At this stage we are left with the challenge of evaluating the master integrals.
Due to their dependence on Mandelstam variables and masses we deem an analytic computation currently unfeasible.
Instead we use the auxiliary mass flow method~\cite{Liu:2017jxz,Liu:2020kpc,Liu:2021wks} for fast numerical evaluation.
We construct a system of differential equations for the master integrals with respect to the squared $W$ boson mass, $m_W^2$.
Using boundary conditions at $m_W^2 \to -i \infty$ we can solve the system at the physical $W$ boson mass.

This approach is very similar to the one employed in earlier work~\cite{Bronnum-Hansen:2020mzk,Bronnum-Hansen:2021olh}.
However, the present situation is complicated by the fact that not all boundary integrals are known analytically or are not known to sufficiently high order in the dimensional regulator, $\epsilon$.
We calculate these boundary integrals numerically by continuation of \emph{internal} $m_t^2$ to the complex plane and taking the limit $m_t^2 \to -i \infty$ for the internal top quark propagators only.

The master integrals for a specific phase space point can now be evaluated by first solving the system of differential equations in (internal) $m_t^2$ and then using this result as boundary condition to solve the system of differential equations in $m_W^2$.

Using this approach we are able to evaluate the 428 master integrals to a precision of 20 digits for the bulk of phase space. Evaluation of a single point for all master integrals takes approximately $30$ minutes on a single core.

\section{Pole structure of the non-factorisable contributions to the scattering amplitude}

Since we only consider non-factorisable contributions, we find that their divergences are remarkably simple.
In fact, there are no divergences of ultraviolet origin and renormalisation amounts to relating the bare and renormalised $\overline{\text{MS}}$ couplings at leading order.
The amplitude does, however, contain poles of infrared origin.

The infrared pole structure for amplitudes at the two-loop level is well understood.
We follow~\cite{Catani:1998bh,Becher:2009qa,Czakon:2014oma} to predict and subtract infra-red poles in order to obtain a finite remainder, $\mathcal{F}$.
This is achieved through the application of a colour-space operator
\begin{align}
\mathcal{A} = \boldsymbol{Z} \mathcal{F}.
\end{align}
The general expression for the infra-red poles
reduces to a remarkably compact expression when the contributions to the non-factorisable part of the amplitude are singled out and interfered with the LO amplitude.
In particular, there are no contributions from collinear divergences and the expansion starts at $\epsilon^{-2}$ instead of $\epsilon^{-4}$.

The pole structure prediction provides an important cross-check of our calculation. Indeed, for the two-loop amplitude we find that the $\epsilon$-poles agree to more than $10$ digits for the bulk of phase space, see Table \ref{tab:poles} for an example.

\begin{table}[!h!tbp]
\begin{center}
\begin{tabular}{l|cc}
                        & $\epsilon^{-2}$ & $\epsilon^{-1}$ \\
                        \hline
                        $\mathcal{A}^{(0)\,\star} \mathcal{A}_{\rm nf}^{(2)}$ & \scriptsize{$-229.094040865466{\color{red}0} -8.978163333241{\color{red}640} i$} & \scriptsize{$-301.18029889447{\color{red}64} -264.17735965295{\color{red}05} i$}  \\ 
                        IR poles & \scriptsize{$-229.0940408654665 -8.978163333241973 i$} & \scriptsize{$-301.1802988944791 -264.1773596529535 i$} \\
                        \hline
                \end{tabular}
                \caption{Comparison of colour-summed interference between the tree-level and non-factorisable, two-loop amplitudes with the infrared pole prediction. The chosen phase space point is given by  $s \approx 104337~\text{GeV}^2$ and $t \approx -5179.68~\text{GeV}^2$. We find agreement to $13$ - $15$ digits.}
                \label{tab:poles}
                \end{center}
\end{table}

\section{Results}

We estimate the impact of the non-factorisable, two-loop contribution on single-top production by computing the double virtual cross section.
It is important to emphasise that this is not a physical cross section as we did not include the correction from real emissions.
However, since we have used a numerical method to evaluate the master integrals, it is important to demonstrate that our results can be used in phenomenological studies.

For the evaluation of the non-factorisable contribution to the cross section we first construct a reliable grid for the evaluation of the LO cross section and simple kinematic distributions.
We randomly draw points from this grid in order to obtain an estimate for the cross section, and find

\begin{align}
\sigma^{ub}_{pp\to dt} = \left ( 90.3 + 0.3 \left ( \frac{\alpha_s(\mu_{\rm nf})}{0.108}  \right )^2  \right ) ~{\rm pb}.
\label{eq:xsvalue}
\end{align}
The first term is the LO cross section, while the second is the non-factorisable NNLO contribution.
We have used the renormalisation scale $\mu = m_t$, but indicate that the scale for the non-factorisable contribution, $\mu_{\rm nf}$, can be varied independently since this contribution first appears at NNLO.
With our scale choice the correction is around $0.3 \%$ and is hence smaller albeit comparable to the factorisable contribution.
A different scale choice can enhance the importance of this contribution (e.g. the transverse momentum of the top quark).

\begin{figure}[t]
    \centering
      \begin{subfigure}{0.32\textwidth}
    \centering
    \includegraphics[width=\textwidth]{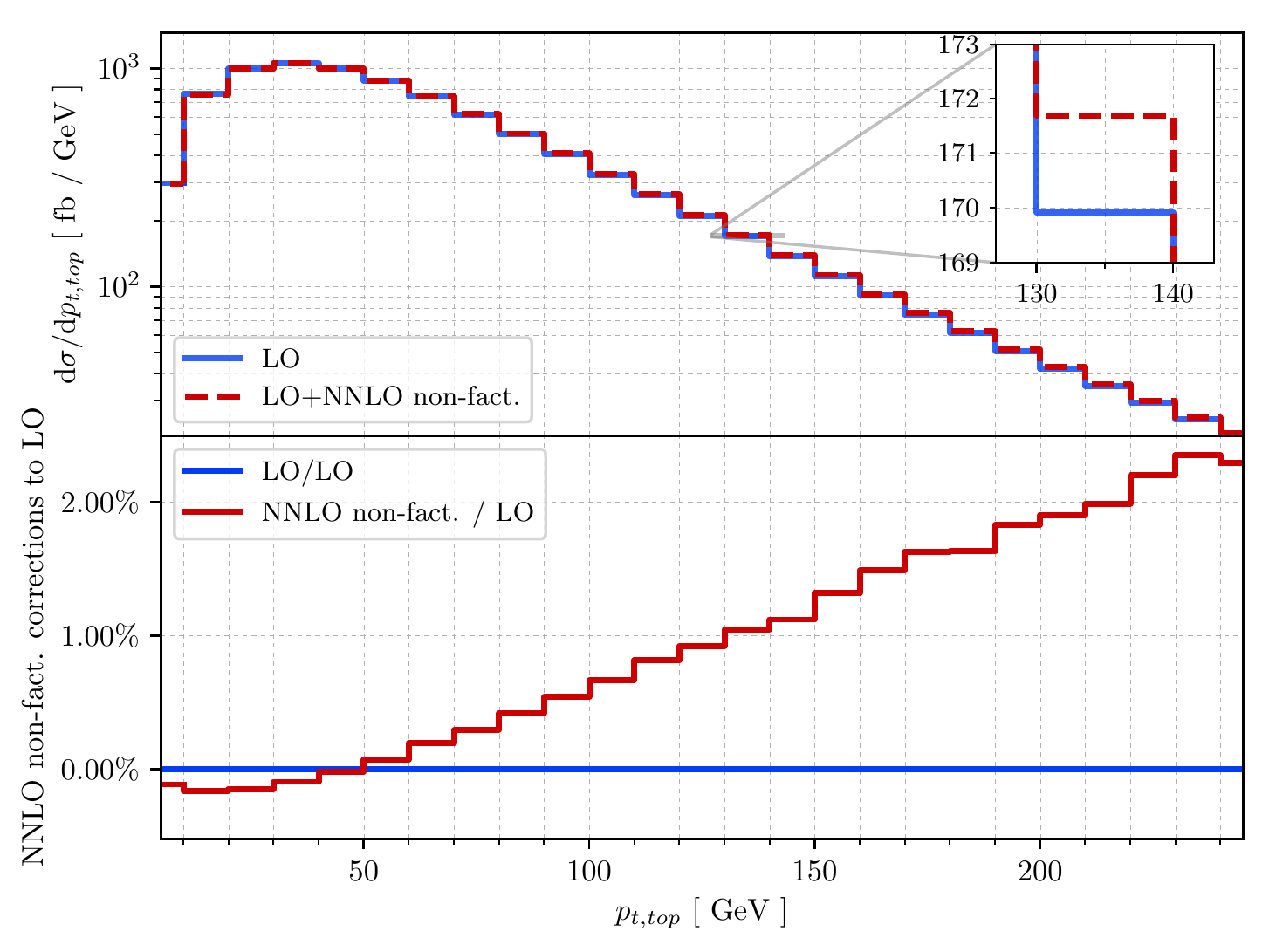}
    \caption{The top-quark rapidity distribution.}
    \label{fig:pttop}
  \end{subfigure}
      \begin{subfigure}{0.32\textwidth}
    \centering
    \includegraphics[width=\textwidth]{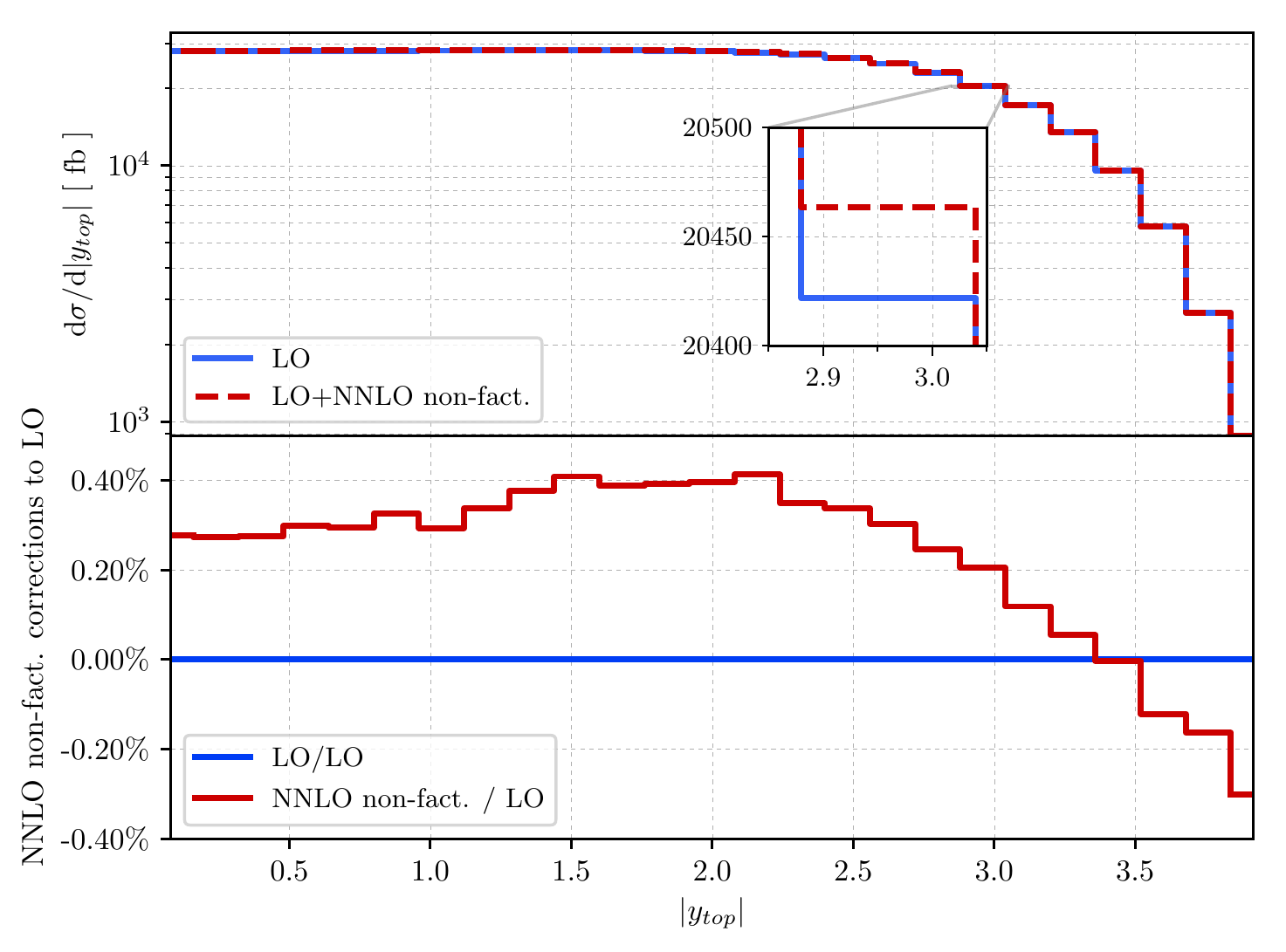}
    \caption{The top-quark rapidity distribution.}
    \label{fig:ytop}
  \end{subfigure}
  \begin{subfigure}{0.32\textwidth}
    \centering
    \includegraphics[width=\textwidth]{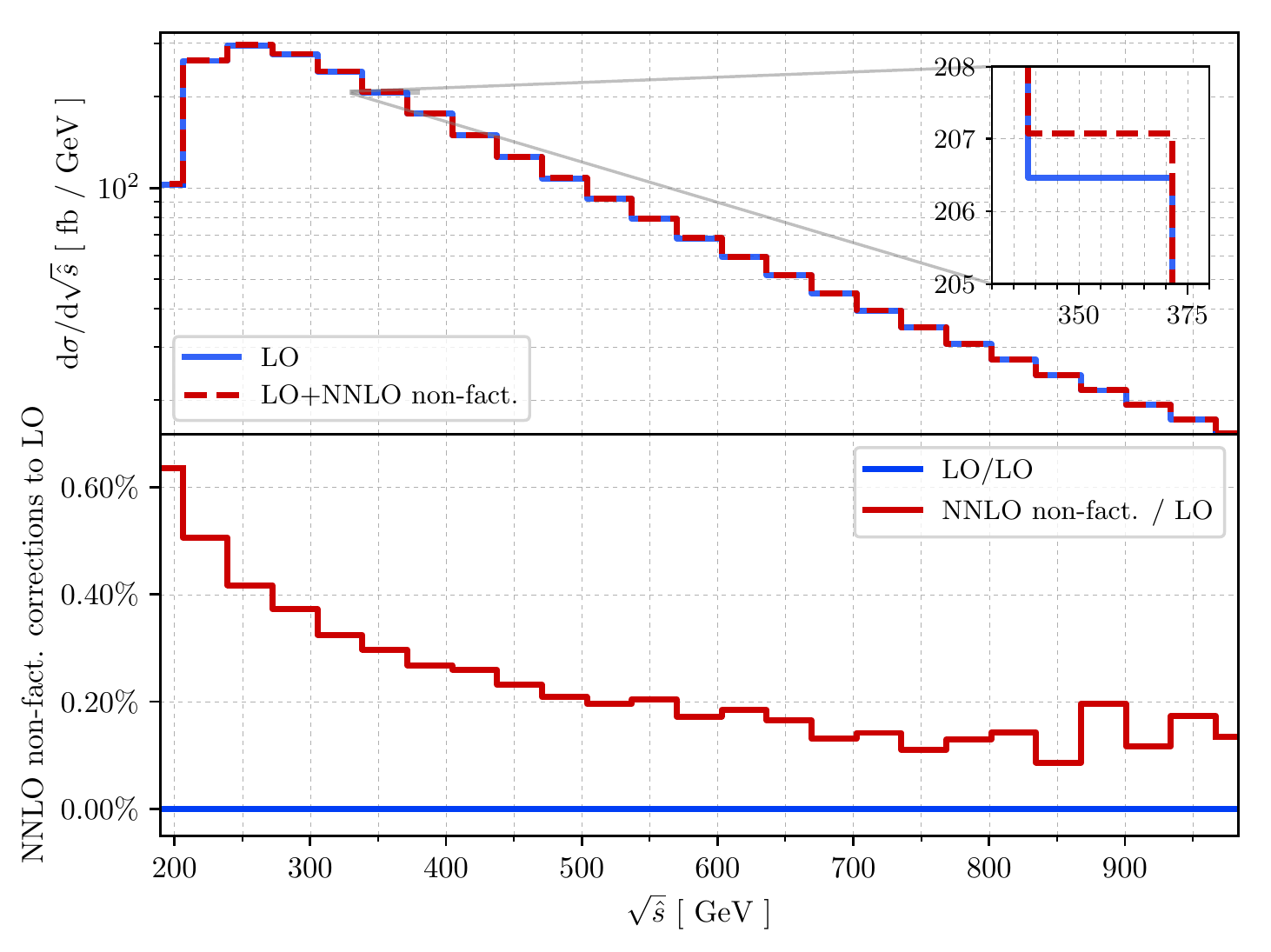}
    \caption{Partonic invariant energy $\sqrt{\hat{s}}$ distribution.}
    \label{fig:shat}
  \end{subfigure}
  \caption{Distributions of the top-quark transverse momentum distribution (left), absolute value of the top-quark rapidity  (middle), and the partonic centre of mass
    energy $\sqrt{\hat{s}}$ (right). Upper panes show leading order distributions as well as distributions with  non-factorisable
    corrections included. 
    Lower panes show the non-factorisable corrections relative to LO distributions. See text for further details.}
  \label{fig:distrib}
\end{figure}

We finish by presenting a few kinematic distributions. In Figure~\ref{fig:pttop} the transverse momentum of the top quark $p_{t,top}$ is plotted and compared to LO.
The correction is small but, interestingly, the factorisable correction has a stronger dependence on the transverse momentum~\cite{Brucherseifer:2014ama,Berger:2016oht,Campbell:2020fhf,Gao:2020ejr} than the non-factorisable.
Hence, the relative importance of these contributions varies across phase space.

In Figure~\ref{fig:ytop}, we compare the impact of the non-factorisable contribution to the LO cross section for the rapidity of the top quark, $y_{top}$.
We show the distribution of the partonic centre-of-mass energy, $\sqrt{\hat{s}}$, in Figure~\ref{fig:shat}.
Also in these distributions the corrections are small, but, as mentioned above, inclusion of real emission contributions is necessary for a complete analysis.

\Acknowledgements
CBH would like to thank the organisers of the 14th International Workshop on Top Quark Physics for the opportunity to present this work.
This research is partially supported by the Deutsche Forschungsgemeinschaft (DFG, German Research Foundation) under grant 396021762 - TRR 257.

\bibliography{eprint}{}
\bibliographystyle{unsrt}
 
\end{document}